# GW170814: Gravitational wave polarization analysis

**Robert C Hilborn**

American Association of Physics Teachers, One Physics Ellipse, College Park, MD 20740, United States of America

**Abstract**

To determine the polarization character of gravitational waves, we use strain data from the GW170814 binary black hole coalescence event detected by the three LIGO-Virgo observatories, extracting the gravitational wave strain signal amplitude ratios directly from those data. Employing a geometric approach that links those ratios to the gravitational wave polarization properties, we find that there is a range of source sky locations, partially overlapping the LIGI-Virgo 90% credible range of GW170814 source locations, for which vector polarization is consistent with the observed amplitude ratios. A Bayesian inference analysis indicates that the GW170814 data cannot rule out vector polarization for gravitational waves. Confirmation of a vector polarization component of gravitational waves would be a sign of post-general relativity physics.

## 1. Introduction

The observation of the binary black hole gravitational wave (GW) event GW170814 by the three LIGO-Virgo observatories [1] allows for a test of the polarization modes of GWs, something not possible with the previous GW observations involving just the two LIGO observatories [2]. General relativity (GR) predicts that GWs will exhibit only tensor polarization modes [1,3-6], an aspect of GR that has not been directly tested before. Any detection of non-tensorial polarization (that is, vector or scalar polarization modes) will be a signal of post-GR physics [7,8]. As C. Will [7] has pointed out, "If distinct evidence were found of any mode other than the two transverse quadrupolar modes of general relativity, the result would be disastrous for general relativity." Hence, determining the polarization properties of GWs is of critical importance in testing GR's description of GWs.

General metric theories of gravity indicate [7,9-13] that there are six possible polarization modes for GWs: two tensor polarization modes (the usual GR prediction), two vector polarization modes, and two scalar polarization modes. It is well known (see, for example, [11-14]) that from the measurements of the signal amplitude ratios among three (or more) observatories, one can determine the polarization properties of the GWs in a way that is independent of the specific model of the binary inspiral dynamics. It turns out, as we shall demonstrate in this paper, that a definitive polarization test requires knowledge of the GW source sky location with a reasonably small range of uncertainty.

This paper builds on three major premises: (1) Given a specified GW source sky location, one can calculate the relative amplitudes for the GW waveform detected by the three LIGO-Virgo observatories based on a purely geometric method. (2) The relative signal amplitudes among the three observatories can be extracted from the GW170814 data with uncertainties small enough to make significant comparisons with the GW polarization predictions. (3) The comparison between the observed and predicted relative amplitudes indicates that both tensor and vector polarization models can account for the relative amplitude observations for a range of source locations consistent with the GW170814 results presented in Ref. [1].

We will use established results [11-15] to show how both the standard tensor polarization modes and "non-standard" vector polarization modes lead to predictions of the ratios of the GW amplitudes seen at the three observatories. (We will not treat the scalar "breathing" and longitudinal polarization modes here.) Vector polarization modes appear in a pure vector theory of gravity [16] and in electromagnetism-like (E&M-like) models of gravitational waves [17,18]. The critical point for our analysis is that by focusing on the relative amplitudes among the three observatories, the analysis becomes purely geometric, depending solely on the mix of polarization modes, the polarization orientation, the GW propagation direction and the orientations of the LIGO-Virgo interferometer observatories [2,5,13].

Comparisons of the observed signal amplitude ratios with GW polarization predictions need to consider the interplay among the GW source sky location (or equivalently, the GW wave propagation direction), the amplitudes and phases of the signals at the three observatories, the wave-front arrival times at the observatories, and the polarization properties of the GWs. The relative amplitudes and phase shifts depend, of course, on the orientation of the observatories' interferometer arms relative to the GW propagation direction, on the mix of polarization modes determined, for a binary orbit source, by the binary's orbital inclination angle, and the GW polarization orientation relative to the detectors.

Binary black hole GW signals can be described in terms of three stages: (1) pre-merger inspiral, (2) merger, and (3) ring-down. We will focus on using the pre-merger inspiral data to analyze the results because we have explicit expressions (see section 3) for the orbital inclination angle and polarization orientation angle dependence of the signals during the binary inspiral. During the merger and ring-down stages, the concomitant strong spacetime curvature means that numerical GR is needed to predict the polarization modes for those phases [19]. There is no fundamental reason for the merger and ring-down mix of polarization modes to be the same as the mix for the inspiral stage. More detailed analyses [20] show that the spins and masses of the orbiting objects determined from the inspiral data match the spins and masses determined from the full numerical GR analysis, giving further justification for focusing on the inspiral stage for the polarization analysis.

The sky location of the GW source can be reasonably well determined from the arrival-time delays of the gravitational wave-front at the three observatories. However, the observed delay times may be affected by phase shifts due to the three detectors' distinct responses to the polarization modes. Generally, those phase shifts are different for the different observatories and are not known a priori [21] since they depend on the binary orbit inclination angle and the polarization orientation angle. In turn, inferences about the polarization of the gravitational waves depend on the source sky location. We explore a range of sky locations and angles of inclination and polarization orientation to study the interdependence of these parameters. It turns out to be important to consider the full range of those angles [22].

Assuming the GW170184 source is a compact binary black hole inspiral and merger (for which there is considerable evidence), we confirm that for the LIGO-Virgo maximum a posteriori GW source sky location, tensor polarization can explain the observed amplitude ratios among the three observatories. This result is not surprising since the LIGO-Virgo analysis that determines the maximum a posteriori source location assumed GW tensor polarization. However, as we shall see, for other source sky locations consistent with the location uncertainty range, vector polarization models also account for the amplitude ratios.

In [1], section VII, the authors state that they considered only GW polarization models that have pure tensor, pure vector or pure scalar polarization modes. The authors note, consistent with the claims in this paper, that the details of the orbital phase model are not important as long as the model gives a good description of the data displayed in Figure 1 of Ref. [1]. Without providing further details, they assert that the GW170814 data favor (in the Bayesian sense) tensor polarization over vector polarization by a factor of 200. (This factor was reduced to 30 in a subsequent re-analysis of the data [23].) However, since the details of those analyses have not yet been published, it is not possible to confirm or contradict those assertions. In section 9, we show that there are good reasons to believe that the more recent Bayes factor [23] is at least an order of magnitude too large.

The strategy in this paper is to expose the detailed dependence of the GW signals on the propagation direction, the orbital inclination angle, the polarization orientation angle, and the detectors' geometry. We shall see that this approach yields results that are consistent with the LIGO-Virgo results [1] under conditions to be described, but it also indicates that the GW170814 data cannot rule out vector polarization for GWs.

## 2. Gravitational wave and detector geometry

In this section we introduce a set of vectors to specify the geometry of the detector orientation, the orbital angle of inclination and the polarization orientation angle, all of which affect the GW signals seen by the LIGO-Virgo observatories.

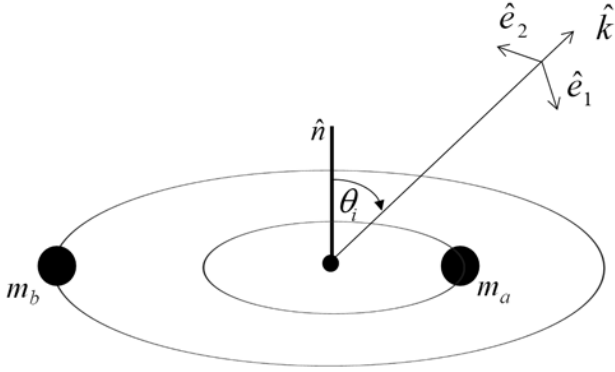

**Figure 1.** A perspective view of the orbits for masses $m_a$ and $m_b$, showing the definition of the angle of inclination $\theta_i$ and unit vectors parallel to and transverse to the gravitational wave propagation direction $\hat{k}$. The unit vector $\hat{n}$ is perpendicular to the plane of the orbit and parallel to the system's orbital angular momentum. The gravitational wave unit vector $\hat{k}$ points from the binaries' center of mass to the observation point. $\hat{e}_1$ lies in the $(\hat{n},\hat{k})$ plane. $\hat{e}_2$ is perpendicular to that plane.

Figure 1 shows the definition of the binary orbit's inclination angle $\theta_i$: the angle between $\hat{k}$, pointing from the source to the observation point, and $\hat{n}$ normal to the orbital plane. The vector $\hat{e}_1$ lies in the $(\hat{n},\hat{k})$ plane while $\hat{e}_2$ is perpendicular to that plane. We shall use $\hat{w} = \{\hat{e}_1, \hat{e}_2, \hat{k}\}$ to refer to the GW unit vectors collectively.

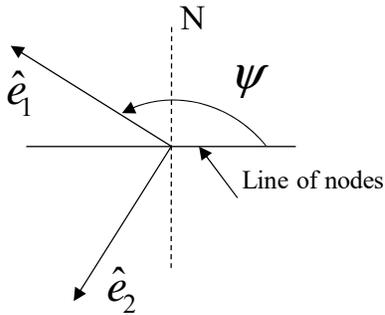

**Figure 2.** The definition of the gravitational wave polarization orientation angle $\psi$. $\hat{k}$ (not shown) points up and out of the page. The Line of nodes is the intersection of the wave front $(\hat{e}_1, \hat{e}_2)$ plane and Earth's equatorial plane. The dashed line is perpendicular to Earth's equatorial plane. N indicates Earth's North Pole.

In this paper, we will focus on the LIGO-Virgo collaboration Hanford (*H*), Livingston (*L*), and Virgo (*V*) observatories, all of which detected the GW170814 event. The detector arms' orientations are specified in terms of unit vectors $\hat{d}_a$ and $\hat{d}_b$, which define the detector plane (with $d = H, L$, or $V$). The local vertical direction is along $\hat{d}_c$. We shall use $\hat{d} = \{\hat{d}_a, \hat{d}_b, \hat{d}_c\}$ to refer to the detector unit vectors collectively. The geometry of the GW observatories is usually specified by the latitude and longitude of the vertex of the interferometer arms and by the unit vectors giving the orientation of the two interferometer arms in the plane perpendicular to the local vertical.

As mentioned previously, we also need to consider the orientation of the GW polarization axes relative to the detector arms. The orientation of the GW polarization may be specified by an angle $\psi$ measured relative to the "Line of nodes" defined by the intersection of the wave front $(\hat{e}_1, \hat{e}_2)$ plane and Earth's equatorial plane. See figure 2.

To calculate the GW detector signal, we also need to know the direction of propagation of the GW. The propagation direction vector that passes through the center of Earth first hits Earth's surface at a point whose location we will specify in terms of latitude and longitude. In section 4, we will discuss the standard method used to find the latitude and longitude of that point for a short GW burst.

Given the orientation of the GW detectors, the GW wave propagation direction, and the polarization orientation, calculating the effects of the GW polarization for the LIGO-Virgo detectors is a straightforward matter of geometry. As mentioned previously, by observing the GW signal amplitude ratios among the three LIGO-Virgo observatories, in principle we need not consider the details of the binary orbit phase evolution. However, for the GW170814 event signal, which is in the detectors' frequency range for only a few tenths of a second and for which the signal amplitude is about the same as the noise amplitude, a model GW waveform is needed to find accurate values of the amplitude ratios. We discuss that issue in section 5.

### 3. Antenna patterns and detector signals

In this section we summarize how to use the vectors introduced in the previous section to find the GW detectors' responses to GWs from orbiting binaries. We will first give the tensor polarization properties associated with standard GR. We then show how those results are modified if we use a GW vector polarization model.

We write each detector's strain signal as the sum of the contributions of two polarization modes (1,2), treating tensor and vector polarization separately [4,8,24]:

$$h_D(t) = h_{D1}(t)\cos 2\Phi(t) + h_{D2}(t)\sin 2\Phi(t), \tag{1}$$

with

$$h_{D1,2}(t) = a(t) f_{1,2}(\theta_i) F_{1,2}(\hat{w}, \hat{d}), \tag{2}$$

where $D = H, L,$ or $V$ and $a(t)$ is a slowly varying amplitude, which depends on the distance from the source to the observation point and on the source properties such as orbital frequency and mass separation. $\Phi(t)$ is the time-dependent orbital phase. The factor of 2 reminds us that the GW frequency is twice the orbital frequency. The details of $a(t)$ and $\Phi(t)$ will be given in section 5. The functions $f_{1,2}(\theta_i)$ give the dependence on the orbital inclination angle, while the antenna pattern functions $F_{1,2}(\hat{w}, \hat{d})$ provide the dependence on the wave propagation direction, the polarization orientation angle, and the detector geometry. In this paper, we assume that the functions $f_{1,2}(\theta_i)$ are time-independent. That means we focus on short GW bursts and, for binary orbits, we assume that spin-orbit and spin-spin effects are negligible [25,26].

For the inspiral stage, we write the overall signal in terms of an amplitude and phase:

$$h_D(t) = A_D(\theta_i, \psi, \hat{w}, \hat{d}, t)\cos(2\Phi(t) + \Phi_D), \tag{3}$$

with

$$A_D = \sqrt{h_{D1}^2 + h_{D2}^2}. \tag{4}$$

The phase angle $\Phi_D$ is given by

$$\Phi_D = \tan^{-1}\left[-\frac{h_{D2}}{h_{D1}}\right]. \tag{5}$$

### 3.1 Tensor polarization

In GR's treatment of gravitational waves [3,4], there are two tensor polarization modes, named "plus" (+) and "cross" (×), with amplitudes $h_{D+}$ and $h_{D\times}$. The antenna pattern functions are [27]

$$F_+(\hat{w},\hat{d}) = \tfrac{1}{2}\left[\left(\hat{e}_1 \cdot \vec{d}_a\right)^2 - \left(\hat{e}_1 \cdot \vec{d}_b\right)^2 - \left(\hat{e}_2 \cdot \vec{d}_a\right)^2 + \left(\hat{e}_2 \cdot \vec{d}_b\right)^2\right]$$
$$F_\times(\hat{w},\hat{d}) = \left(\hat{e}_1 \cdot \vec{d}_a\right)\left(\hat{e}_2 \cdot \vec{d}_a\right) - \left(\hat{e}_1 \cdot \vec{d}_b\right)\left(\hat{e}_2 \cdot \vec{d}_b\right).$$
(6)

The inclination angle dependence for a binary orbit GW source is given by [12,25]

$$f_+(\theta_i) = (1 + \cos^2 \theta_i)/2$$
$$f_\times(\theta_i) = -\cos \theta_i .$$
(7)

The antenna pattern dependence on the polarization orientation angle is via the GW vectors $\hat{e}_1$ and $\hat{e}_2$. For a periodic GW source, there is a 90° temporal phase difference between the two tensor polarization modes. See Eq. (1).

*3.2 Vector polarization*

As mentioned previously, in generalized metric theories of gravity, GWs may have a mix of tensor, vector, and scalar polarization modes. In a pure vector theory of gravity [16] and in E&M-like (vector) models of GWs [17,18], the GW waves have two vector polarization components (exactly like the description of polarization for E&M waves), traditionally labeled "*x*" and "*y*". These correspond to linear polarization along $\hat{e}_1$ and $\hat{e}_2$, respectively. For a simply periodic wave source, the two vector polarization modes have a temporal phase difference of 90°. See Eq. (1).

The vector model antenna pattern functions are [27]

$$F_x(\hat{w},\hat{d}) = \left(\hat{k} \cdot \vec{d}_a\right)\left(\vec{d}_a \cdot \hat{e}_1\right) - \left(\hat{k} \cdot \vec{d}_b\right)\left(\vec{d}_b \cdot \hat{e}_1\right)$$
$$F_y(\hat{w},\hat{d}) = \left(\hat{k} \cdot \vec{d}_b\right)\left(\vec{d}_b \cdot \hat{e}_2\right) - \left(\hat{k} \cdot \vec{d}_a\right)\left(\vec{d}_a \cdot \hat{e}_2\right).$$
(8)

In a vector theory of gravity [16] and in E&M-like models [17,18] of GWs, the pre-merger dependences on the angle of inclination for a binary orbit GW source are

$$f_x(\theta_i) = \sin 2\theta_i$$
$$f_y(\theta_i) = -2\sin \theta_i .$$
(9)

It is important to note that the antenna pattern functions in Eqs. (6) and (8) will hold for any gravitational wave theory since they are purely geometric expressions of what we mean by tensor polarization and vector polarization, respectively.

*3.3 Amplitude ratios*

We are interested in the ratios of amplitudes for the different observatories, for example $A_V/A_L$, and phase differences, for example $\Phi_V - \Phi_L$, as functions of the angles $\theta_i$ and $\psi$ for a specified

GW source sky location. By examining Eqs. (1), (2) and (3), we see that using those ratios removes the effects of the time-varying orbital phase $\Phi(t)$ and amplitude $a(t)$. Comparison with the GW170814 data will then in principle give us information about possible values of $\theta_i$ and $\psi$ consistent with those ratios.

To calculate the various scalar products in Eqs. (6) and (8), it is helpful to express each of the vectors in terms of an Earth-fixed coordinate system with $\vec{z}_E$ pointing from Earth's center towards the North pole, $\vec{x}_E$ running from the center of Earth along the prime meridian (which intersects the surface at 0º longitude) and $\vec{y}_E$ forming a right-handed Cartesian coordinate system with $\vec{x}_E$ and $\vec{z}_E$. The components of each of the detector unit vectors in such an Earth-fixed system are available from the LIGO-Virgo collaboration [28,29]. The gravitational wave unit vectors can be expressed in terms of the Earth-fixed unit vectors by using the source latitude and longitude at the time of the event (or equivalently, the source's Right Ascension and declination) and the unknown polarization orientation angle $\psi$. For an alternative, but equivalent, approach to dealing with GW and detector geometry, see [30,31].

## 4. GW source sky location

We now review how to use the observed wave-front arrival-time delays among the three observatories to determine the celestial coordinates of the gravitational wave source. Those results will define the appropriate range of source sky locations to be used in the subsequent analysis (sections 6-9). We assume that the GW signal duration is short (in practice from a few tenths of a second to a few tens of seconds for recent observations), so the source sky location relative to the rotating and orbiting Earth does not change significantly during the GW observation time.

We denote the position of the Hanford observatory relative to the Livingston observatory as $\vec{r}_{LH}$ and the analogous position of Virgo as $\vec{r}_{LV}$. Given those vectors, it is easy to see that the arrival-time delays are given by

$$\Delta t_{LH} = \hat{k}\left(\theta_{lat}, \phi_{long}\right) \cdot \vec{r}_{LH} / c \tag{10}$$

$$\Delta t_{LV} = \hat{k}\left(\theta_{lat}, \phi_{long}\right) \cdot \vec{r}_{LV} / c, \tag{11}$$

where $\theta_{lat}$ and $\phi_{long}$ are the latitude and longitude of the GW source at the time of the event. The gravitational wave speed is assumed to be the usual speed of light. The GW propagation unit vector can be expressed in the Earth-fixed coordinate system as

$$\hat{k}(\theta_{lat},\phi_{long}) = -\cos\theta_{lat}\cos\phi_{long}\ \hat{x}_E + \cos\theta_{lat}\sin\phi_{long}\ \hat{y}_E - \sin\theta_{lat}\ \hat{z}_E\ . \tag{12}$$

Eqs. (10) and (11) provide two equations for the two unknown angles $\theta_{lat}$ and $\phi_{long}$.

We will now deploy the expressions given above to extract information about the GW source location based on the observations of GW170814 described in [1]. The reported gravitational wave-front arrival-time delays are $\Delta t_{LH} = 8\ \pm1\,\text{ms}$ and $\Delta t_{LV} = 14\ \pm1\,\text{ms}$, for Livingston-Hanford and Livingston-Virgo respectively. ( [1] does not explicitly state the uncertainties in the arrival-time delays, but examination of the data shown in figure 4 below indicates that 1 ms is a reasonable estimate of that uncertainty.) We note that observed time-delays are also affected by possible intrinsic phase shifts associated with the mix of polarization modes. We discuss those phase shifts in section 8.

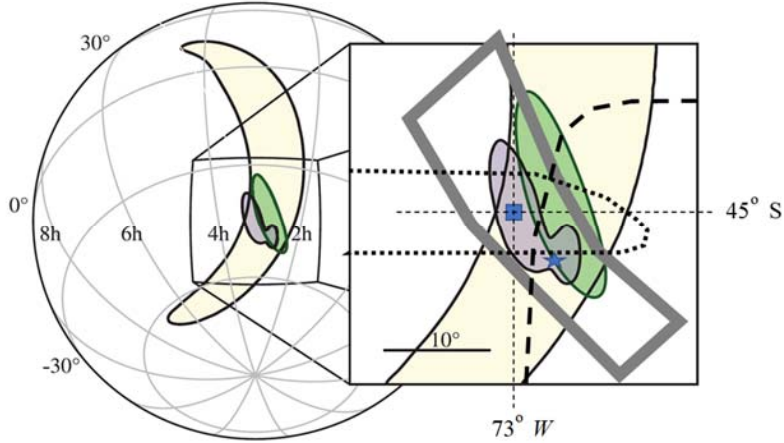

**Figure 3.** Based on Figure 3 CC BY 3.0 of [1], this figure shows the LIGO-Virgo determination of the range of possible locations of the GW170814 source in equatorial (celestial) coordinates. The inset is a gnomonic projection. The thin solid dark contours in the inset represent the LIGO-Virgo 90% credible regions. The light (yellow) shading is the localization using only the two LIGO sites. The intermediate (green) shading is the rapid localization results using data from all three observatories, while the smaller asymmetric range is the full parameter estimation localization [1]. The inset's filled square indicates approximately the maximum a posteriori source location: RA $03^\text{h}\ 11^\text{m}$, dec $-44°57^m$. The star symbol shows the source location

$\theta_{lat} = 50°\text{S}$, $\phi_{long} = 77°\text{W}$ calculated from the observed arrival-time delays. The light gray lines bound the range of source locations consistent with the range of uncertainty for the arrival-time delays. The dark dashed line indicates the approximate range of source locations, discussed later, for which a vector polarization model can explain the relative amplitude observations. The dotted line outlines the analogous region for a tensor polarization model. All boundaries are approximate.

Using those time delays and the relative position vectors $\vec{r}_{LH}$ and $\vec{r}_{LV}$, we find the source latitude and longitude at the time of the event to be $\theta_{lat} = 50°$ S and $\phi_{long} = 77°\text{W}$ (to two significant figures). The location uncertainties corresponding to the 1 ms arrival-time delay uncertainty are roughly 10° in latitude and 4°-5° in longitude, dominated by the Livingston-Hanford time-delay uncertainty. Note that this method of determining the source location is a purely geometric one, independent of GW polarization and details of the GW waveform. The region of source locations constrained by the arrival-time delays and their uncertainties is bounded by the light gray lines in figure 3.

The full parameter Bayesian estimation of [1] constrains the source location to a 90% credible area of 60 deg² with a maximum a posteriori position of right ascension RA = $03^h11^m$ and declination = $-44°57^m$ (J2000). The GW170814 Fact Sheet [32] states that the maximum a posteriori location is equivalent to $\theta_{lat} = 45°$ S and $\phi_{long} = 73°\text{W}$. Within the uncertainties, the latitude and longitude results found in this paper from the arrival-time delays agree with the GW170814 maximum a posteriori source location mentioned previously. Note that the largest possible value of the wave front arrival-time delay $t_{LH}$ is 10 ms, which occurs when the GW travels (with speed $c$) parallel to the line joining the $L$ and $H$ detector vertices. This occurs for sky locations in the upper-left sector of the inset in figure 3.

The region in figure 3 to the right of and below the dark dashed line is the area in which the vector polarization model is consistent with the relative amplitudes of the observatories' GW strain signals. The analogous region for the tensor polarization model is inside the dark dotted line. In both cases the regions extend beyond the boundaries of the gnomonic projection.

The main point of this section is that the GW source location parameters are sensitive to the arrival-time delays among the three observatories. The recently observed binary neutron star event

GW170817 [33] is a special case because the source's sky location could be determined rather precisely from independent observations of electromagnetic emissions from the colliding neutron stars [34]. As discussed elsewhere [35], the precise source sky location for GW170817 allows for a definitive test of GW polarization.

## 5. Extracting GW signal amplitude ratios

In section 3, we recognized that knowing the GW signal amplitude ratios among the three LIGO-Virgo observatories is sufficient, under favorable circumstances, to make definitive statements about the polarization properties of the GWs. Because those signal amplitude ratios play such an important role in the polarization analysis, we have used three distinct methods to extract those ratios from the LIGO-Virgo interferometers' GW170814 strain data: (1) a direct nonlinear least-squares fit of the expected GW "chirp" signal to the data, (2) an integrated signal method [35] that is closely related to a "matched filter" analysis of the data, and (3) a Fourier-amplitude maximum-likelihood method [36] that is a specialized instantiation of the Bayesian method used by the LIGO-Virgo collaboration [36] in its full-parameter fit of GW data. Our only assumption is that the relative amplitudes and phases of the polarization modes do not change significantly during the binary inspiral. We find that all three methods give consistent results within the ranges of uncertainty.

The extraction of reliable signal amplitude information for the GW170814 event is challenging because the GW signal amplitude is of the same order of magnitude as the noise amplitude in the frequency range of the LIGO-Virgo observatories. Furthermore, as mentioned previously, the GW170814 signal is within that frequency range for only a few tenths of a second, making the application of noise-averaging techniques more difficult. Nevertheless, a careful examination of the GW170814 data provides a useful case study of GW polarization tests.

*5.1 The GW chirp signal*

As is well known [3,4], in the early inspiral stage of a binary orbit GW event, the GW waveform can be accurately described by a relatively simple chirp signal. The formulation of the chirp signal is simplified if we introduce the so-called chirp mass $M_{\text{chirp}} = M\eta^{3/5}$, where $\eta = m_a m_b / (m_a + m_b)^2$ is a dimensionless mass ratio and $M = m_a + m_b$ is the total mass of the binary system. To set the distance scale, we use $R_{\text{S-chirp}} = 2GM_{\text{chirp}}/c^2$, where $G$ is the usual Newtonian gravitational constant. To set the time scale, we use $\tau_{\text{S-chirp}} = R_{\text{S-chirp}}/c$. ($R_S$ is the Schwarzschild radius for a mass $M$, but we will not need that physical interpretation in what follows.)

It is also well known [4,37,38] that $a(t)$ and $\Phi(t)$ in Eqs. (2) and (3) are the same for all of the detectors (once the arrival-time delays are taken into account). In the simplest post-Newtonian approximation, $a(t)$ is a slowly varying overall amplitude given by

$$a(t) = \left(\frac{2}{5}\right)^{3/4} \frac{1}{\eta^{1/5}} \frac{R_{\text{S-chirp}}}{R} \left(\frac{t_c - t}{\tau_{\text{S-chirp}}}\right)^{-1/4}, \quad (13)$$

where $t_c$ is the so-called coalescence time at which, in this model, the mass separation goes to 0 and the orbital frequency becomes infinite. $R$ is the distance between the binary orbit and the detector.

The time-dependent orbital angular frequency $\omega$ is expressed as

$$\omega(t) = \left(\frac{5}{2}\right)^{3/8} \frac{1}{4\tau_{\text{S-chirp}}} \left[\frac{t_c - t}{\tau_{\text{S-chirp}}}\right]^{-3/8}, \quad (14)$$

and the accumulated orbital phase $\Phi(t)$ is given by

$$\Phi(t) = \int_{t_0}^{t} \omega(\tau)\,d\tau = -\left(\frac{2}{5}\right)^{5/8} \left[\frac{t_c - t}{\tau_{\text{S-chirp}}}\right]^{5/8} + \Phi_c, \quad (15)$$

where $\Phi_c$ is the phase at the coalescence time. We note that the results given in Eqs. (13)–(15) are the same for both lowest-order GR and Svidzinsky's vector theory of gravity [16].

In the numerical results that follow, we use the chirp mass in the detector frame (compared to the chirp mass in the source frame, as reported by the LIGO-Virgo Collaboration). For the GW170814 event, the detector-frame chirp mass is $M_{\text{chirp}} = 26.9^{+1.4}_{-1.1}$ M$_\odot$ (about 10% larger than the source-frame value [1]). M$_\odot$ is the solar mass.

We will now use the three methods to find the relative amplitudes among the three LIGO-Virgo observatories for the GW170814 event, assuming that Eqs. (3) and (13)–(15) provide an accurate description of the GW inspiral waveform. We employ the LIGO-Virgo "strain data after noise subtraction" (sampling rate 16384 Hz) available from the LIGO-Virgo Gravitational Wave Open Science Center website [39]. We select a four-second-long time series centered on the GW170814 event. Those data are windowed with a Tukey window function with 0.5 s turn-on and turn-off times and time-shifted to remove the arrival-time delays described previously. The windowed data are then subject to a 40-300 Hz band-pass filter to remove both low-frequency and high-frequency noise. For the Virgo data, we also apply 50 Hz, 62-63 Hz, and 150 Hz notch filters to remove the large amplitude power mains and mechanical noise at those frequencies. The time series reconstructed from the filtered data are used in the following analysis. Since we want to concentrate on the inspiral phase of the binary orbit, we select data for the detailed analysis with $t \in [t_p - 0.1 \text{ s}, t_p - 0.005 \text{ s}]$, where $t_p$ is the time at which the GW peak occurs. The details are given below for each of the three analysis methods. All three methods were calibrated by injecting simulated chirp signals into LIGO-Virgo noise time series data beginning a few seconds after the GW170817 event.

## 5.2 Nonlinear least-squares fit

For the first method, we use a nonlinear least-squares fit of Eq. (3), the theoretical GW waveform, to the actual data. The results shown below include an overall scaling to put the numerical values in a range convenient for plotting. (According to figure 1 of [1], the peak strain of the Livingston GW signal is about $0.5 \times 10^{-21}$.) For the fit we use data from $t = 0$ to $t = 0.065 \text{ s}$, as shown in figure 4.

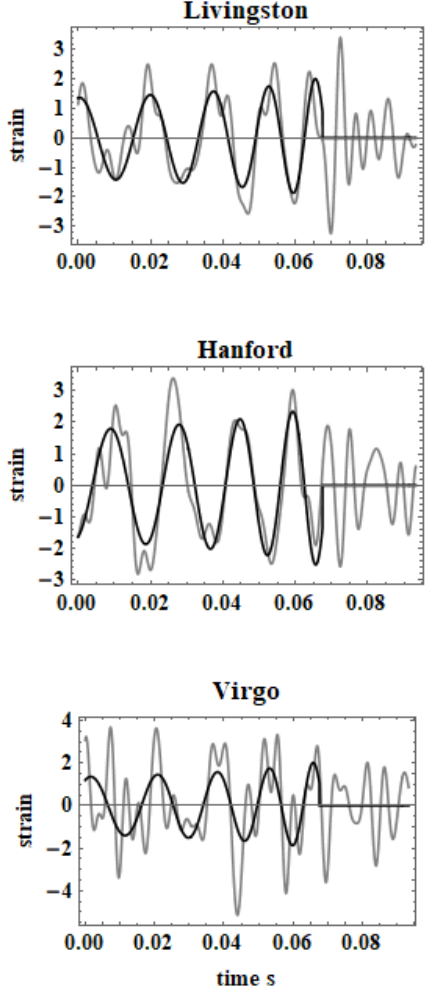

**Figure 4.** Plots of the LIGO-Virgo GW170814 interferometer strain signals (light curves), with signal processing as described in the text, along with the results (dark curves) of a nonlinear least-squares fit of the theoretical GW waveform as functions of time (s). The strain amplitude scale is arbitrary but the same for all three observatories' data. The plotted strain signals extend beyond the merger (at $t_p \approx 0.07$ s) and ring-down. The data region for the fit ends at $t \approx 0.065$ s.

The nonlinear fit routine adjusted the overall amplitude of the strain signal, the coalescence time $t_c$ and the phase $\Phi_D$ for each of the detectors to minimize the sum of the squares of the deviation between the data and the theoretical chirp signal. Ideally, the analysis yields the same $t_c$ for all three detectors if the correct arrival-time delay shifts have been applied to the data. We found those values were the same within a relative uncertainty of ±0.4%. As described in section 3, the intrinsic phases $\Phi_D$ depend on the GW polarization and the orientation of the detector arms relative to the GW propagation direction.

The nonlinear least-squares fits gave the following amplitude ratios:

$$\left|\frac{H}{L}\right| = 1.26 \pm 0.06 \qquad \left|\frac{V}{L}\right| = 1.04 \pm 0.08 , \tag{16}$$

where the (approximately one sigma) uncertainties were estimated by varying the length of the time series used in the fit. (The uncertainties associated with the fit itself are considerably smaller.)

*5.3  Accumulated signal method*

The second method we employ to extract the relative GW signal amplitudes is the "accumulated signal" method devised by A. Svidzinsky [35]. In this method, we integrate the detectors' strain signals $s(t)$ with the exponential of the orbital phase $\Phi(t)$ as given in Eq. (15) to yield an accumulated signal $A_S(t)$:

$$A_S(t) = \int_{t_0}^{t} (t_c - \tau)^{1/4} s(\tau) e^{-i2\Phi(\tau)} d\tau. \tag{17}$$

The integrand factor $(t_c - \tau)^{1/4}$ is included to remove the gradually increasing amplitude of the chirp waveform. It is easy to see that if the detector signal and the GW phase $2\Phi(t)$ have the same functional form, the accumulated signal will increase linearly with $t$. (We may safely ignore a small amplitude term that oscillates at $4 d\Phi/dt$.) The contribution of (zero-mean) noise in the detector signal will average to zero over sufficiently long time periods. Plotting $A_S(t)$ as a function of time should then yield a straight line whose slope is proportional to the signal amplitude. Figure 5 shows the results of this analysis, using $t_c$ and $\Phi_D$ determined from the nonlinear least-squares fit of section 5. 2. Note that $t_0$ in Eq. (17) corresponds to $t = 0$ in figure 4.

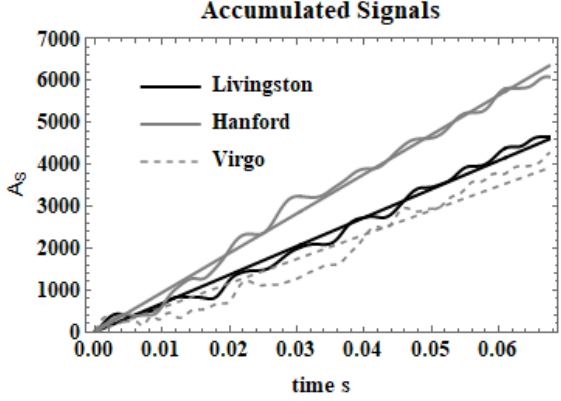

**Figure 5.** The accumulated signals $A_S(t)$ for the three LIGO-Virgo detectors' GW170814 strain signals plotted as a function of accumulation time along with the linear least-square fits (straight lines) to those data. The numerical scale for $A_S(t)$ is arbitrary but the same for all three detectors.

From figure 5, we see that the accumulated signals for the three detectors are reasonably approximated by straight lines, although the data for the Virgo detector are relatively noisy. Using the slopes of those lines as determined by a least-squares fit, we find that the amplitude ratios are given by

$$\left|\frac{H}{L}\right| = 1.38 \pm 0.08 \quad \left|\frac{V}{L}\right| = 0.93 \pm 0.15, \qquad (18)$$

where the uncertainties were estimated by varying the integration limits $t_0$ and $t$.

### 5.4  *Fourier amplitude maximum-likelihood method*

The third method of estimating the relative signal amplitudes is based on the analysis used by the LIGO-Virgo collaboration to estimate the binary source and GW parameters from the GW data [36,40,41]. The key concept is the likelihood [42,43] of the GW signal data given the theoretical model and its parameters, denoted collectively by $\Theta$. The analysis is carried out in Fourier space to take into account the frequency-dependent noise in the GW interferometers. The likelihood $L(s|\Theta)$ of the strain data $s$ given the parameters $\Theta$ is expressed as [44]

$$L(s|\Theta) = \sum_{j=1}^{N_s} \frac{1}{\sqrt{2\pi\sigma_j^2}} \exp\left[-\frac{|\tilde{s}(f_j) - \tilde{h}(f_j)|^2}{2\sigma_j^2}\right], \qquad (19)$$

where $\sigma_j$ is the frequency-dependent noise amplitude spectral density at frequency $f_j$, $\tilde{s}(f_j)$ is the Fourier amplitude of the strain data (GW signal plus noise), $\tilde{h}(f_j)$ is the Fourier amplitude of the theoretical GW waveform, and $N_s$ is the number of samples in the frequency spectrum. Eq. (19) assumes that the noise signal has a Gaussian distribution with zero mean. For our purposes, we restrict the frequencies to lie between 50 and 130 Hz to avoid excessive low frequency noise (particularly in the Virgo data) and to stay within the range of validity of the waveform given by Eq. (3). In practice, it is more convenient to calculate the negative of the natural logarithm of the likelihood function:

$$N_L(s|\Theta) = \frac{1}{2}\sum_{j=1}^{N_s} \frac{\left|\tilde{s}(f_j) - \tilde{h}(f_j)\right|^2}{2\sigma_j^2} + \frac{1}{2}\log\left[2\pi\sigma_j^2\right]. \tag{20}$$

The square of the amplitude spectral density is estimated by using the Welch method [36,45] applied to 10 s of each detector's strain data starting 5 s after the GW170814 event. To be consistent with the signal processing of the GW event time series, we apply a Tukey window with 0.5 s rise and fall times to the data. Each of the windowed time series is then subject to the 40-300 Hz band-pass filter used for the GW event time series. The previously described notch filters are applied to the Virgo noise data.

Each of the noise time series, reconstituted using the inverse Fourier transform, is divided into overlapping time segments, whose lengths match the length of the GW event time series used in the likelihood analysis. Each time segment is Tukey windowed and then Fourier transformed. The absolute-value-squared of the Fourier amplitudes of the segments are averaged over the segment population. Those averages are proportional to the noise power spectral density for each detector.

To extract the maximum-likelihood value of the signal amplitude for the theoretical model, we find the value of the amplitude of $h(t)$ that minimizes $N_L(s|\Theta)$, which, of course, is equivalent to maximizing the likelihood function with respect to that amplitude. In evaluating the waveform model's Fourier amplitudes, we select a segment of the predicted time series data (described previously) and apply a Tukey window with 0.05 s rise and fall times to avoid "edge effects" in the Fourier spectrum of the time series. The windowed theoretical GW waveform is then Fourier transformed and subject to the 40-300 Hz band-pass filter. It is then straightforward to carry out the sum in Eq. (20) and find the amplitude that minimizes the result. The results for the LIGO-Livingston data are shown in figure 6.

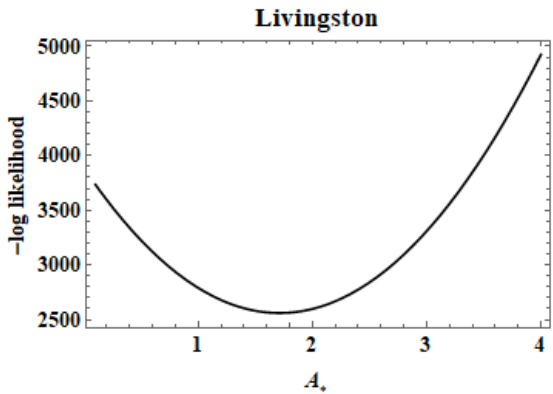

**Figure 6.** A plot of the negative logarithm of the likelihood function [Eq. (20)] for the Livingston GW170814 data, selected as described in the text, as a function of the amplitude $A_*$. The second term in the sum in Eq. (20) is independent of the amplitude of $h(t)$ and has been ignored in this calculation.

Figure 6 shows a plot of $N_L(s|\Theta)$ for the Livingston GW170814 data with the sum in Eq. (20) evaluated over frequencies between 50 and 130 Hz. We used the values of $t_c$ and $\Phi_D$ from the section 5.2. nonlinear least-squares fit, though in principle, we could also use the maximum likelihood method itself to determine their values. For the Livingston data, the minimum of $N_L(s|\Theta)$ as a function of amplitude occurs at $A_* = 1.71$. The absolute value of the amplitude depends on the scaling used for the data and the theoretical model; however, that scaling is the same for all three detectors' likelihood functions.

After carrying out the same analysis for the Hanford and Virgo data, we find the amplitude ratios:

$$\left|\frac{H}{L}\right| = 1.14 \pm 0.08 \quad \left|\frac{V}{L}\right| = 0.84 \pm 0.10 \;, \tag{21}$$

where the (approximately one sigma) uncertainties were estimated by varying the frequency range included in the likelihood function. Note that an overall scaling of the noise power spectral density does not affect the determination of the amplitude ratios.

*5.5  Summary of the amplitude ratio data*

Calculating the weighted average of the results from sections 5.2., 5.3., and 5.4. yields estimates for the amplitude ratios to be used in subsequent sections to draw conclusions about the GW polarization. We find that the weighted averages and their (approximately one sigma) uncertainties are given by

$$\left|\frac{H}{L}\right| = 1.26 \pm 0.07 \qquad \left|\frac{V}{L}\right| = 0.96 \pm 0.1. \tag{22}$$

Ref. [1] does not cite the GW170814 signal amplitudes explicitly. Figure 1 of [1] displays the "whitened" GW signals; that is, GW strain signals reconstituted from the Fourier transform of the raw strain signal, after band-passing and dividing each Fourier amplitude of the signal by the corresponding Fourier amplitude of the noise strain amplitude spectral density. As a result, the reconstructed waveforms for the three observatories look somewhat different because the whitening emphasizes different frequency bands in the three detectors' signals. To extract the actual ("unwhitened") waveforms, the process must be reversed using the noise amplitude spectral density displayed in figure 2 of [1]. The estimated "unwhitened" results are $|H/L| \approx 1 \pm 0.2$ and $|V/L| \approx 0.8 \pm 0.3$, in rough agreement with the results stated in Eq. (22).

The LIGO-Virgo *GW170814 Fact Sheet* [32] states that the Hanford:Livingston:Virgo peak GW strain amplitudes are $6:6:5 \times 10^{-22}$, which gives relative amplitude ratios different from those in Eq. (22). (The *Fact Sheet* does not provide any amplitude uncertainty information.) As we discuss in section 10, there are good physics reasons to expect that the inspiral relative amplitude ratios might be different from the peak (merger and ring-down) ratios.

**6.  Polarization orientation, orbital inclination and the relative strain amplitudes**

We now show how to use the GW amplitude ratios from the previous section along with specified source sky locations to determine what limits (if any) can be placed on the GW binary source orbital inclination angle $\theta_i$ and the polarization orientation angle $\psi$. We will examine the predicted GW signal amplitude ratios (Hanford/Livingston and Virgo/Livingston) and the corresponding signal phase differences as a function of those angles and see, for pure tensor and pure vector polarization models what ranges are consistent with the relative signal amplitude observations.

A few reminders before we describe the details: Eqs. (1) and (2) are restricted to the pre-merger GW signals emitted before the two black holes coalesce. We expect that the polarization properties of the GWs might change during the merger and ring-down phases due to the strong spacetime curvature produced during those phases [46]. Extensions to the merger and ring-down phases are discussed briefly in section 10.

Also, Eqs. (1) and (2) do not take into account possible spin-orbit and spin-spin effects if the black holes are rotating [25]. These effects will cause the orbital angular momentum direction (which is, of course, perpendicular to the plane of the binary orbits) to precess about the total angular momentum direction. This precession leads to a modulation in the ratio of the polarization mode amplitudes along the observation direction and consequently a modulation of the signal amplitudes at the detectors. Furthermore, the total angular momentum of the binary system will change because the GWs themselves carry away angular momentum. For simplicity's sake we will ignore those complications in this analysis. Ref. [1] indicates that any spin-orbit precession effects, if present, are small for GW170814. A treatment of GWs emitted by binary black holes with spin is given in [26,38].

We will start with the LIGO-Virgo GW170814 maximum a posteriori source sky location: $\theta_{lat} = 45°$ S and $\phi_{long} = 73°$ W and examine both the standard GR (tensor polarization) and vector polarization predictions for the relative amplitudes. Later we will explore how the results vary when the source location changes.

### 6.1 Tensor polarization model

First, we consider the tensor polarization prediction for the Livingston observatory signal amplitude as a function of the orbital inclination angle $\theta_i$ and the polarization orientation angle $\psi$. This analysis will help build our intuition about what to expect for the ratios of amplitudes among the various observatories.

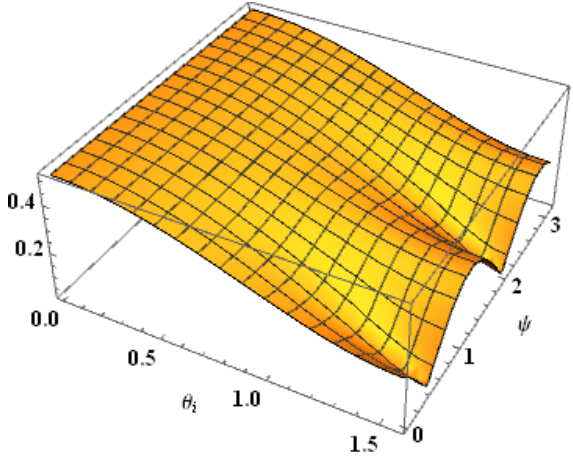

**Figure 7.** The tensor polarization prediction for the Livingston observatory strain signal amplitude (arbitrary units) as a function of the orbital inclination angle $\theta_i$ and the polarization orientation angle $\psi$, both in radians. The amplitude is symmetric about the $\theta_i = \pi/2$ line (not shown). The assumed source sky location is $\theta_{lat} = 45°$ S and $\phi_{long} = 73°$ W.

For small angles of inclination (that is, the observation direction is nearly perpendicular to the plane of the binary orbit), the tensor polarization model (see Eq. (7)) predicts equal amplitudes for the cross and plus polarization modes. Since the temporal phase difference between the modes is $\pi/2$, we have the equivalent of "circular polarization," for which the GW strain amplitude is independent of the polarization angle $\psi$ as illustrated in figure 7.

As $\theta_i \to \pi/2$, the cross polarization mode amplitude $h_\times$ goes to zero, leaving only plus polarization. We see from figure 7 that when $\theta_i \to \pi/2$ the amplitude has a strong dependence on $\psi$ with a periodicity of $\pi/2$, characteristic of the tensor modes of quadrupole radiation. (For vector polarization, the periodicity is $\pi$.) The amplitude graphs for the Hanford and Virgo signals are similar. Also note that the maximum amplitude occurs for $\theta_i$ near 0 and $\pi$, so there may be an observational bias [47] for "face-on" binary orbits. That is, if GR's tensor polarization model is correct, for binaries with identical masses at identical distances, the observations will favor those orbits that are face-on versus those that are "edge-on" ($\theta_i \approx \pi/2$).

We now turn our attention to the amplitude ratios among the different observatories since those ratios are expected to be independent of the time-dependent frequency and amplitude of the GW, at least during the inspiral phase.

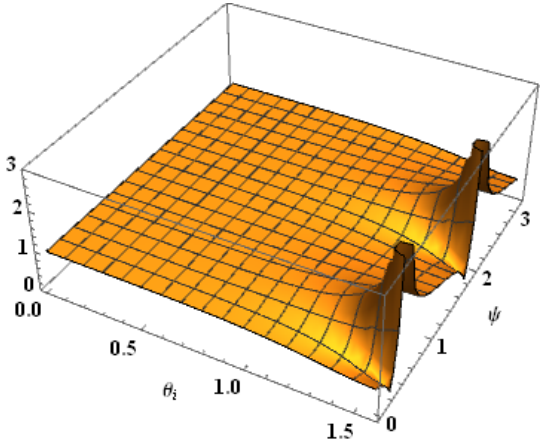

**Figure 8.** A plot of the Hanford-to-Livingston GW tensor polarization signal amplitude ratio as a function of the orbital inclination angle $\theta_i$ and the polarization orientation angle $\psi$, both in radians. The "spikes" near $\theta_i \approx \pi/2$ occur in regions of $\psi$ for which the Livingston amplitude is very small. Source sky location: 45° S, 73° W.

To get a sense of how the amplitude ratios depend on the orbital angle of inclination and the polarization orientation angle, we plot in figure 8 the tensor polarization model predictions for the |H/L| ratio. We see that for $0 \le \theta_i \le 1$ the ratio is about 1, smaller than the |H/L| ratio deduced from the data in section 5. However, for $\theta_i \to \pi/2$, where the plus polarization mode dominates, we find ratios that are strongly dependent on the polarization orientation angle $\psi$. This result contradicts the common statement [1,5] that because of the near-alignment of the Livingston and Hanford observatory interferometer arms, the relative signals at those two observatories are not sensitive to the polarization properties of the GWs.

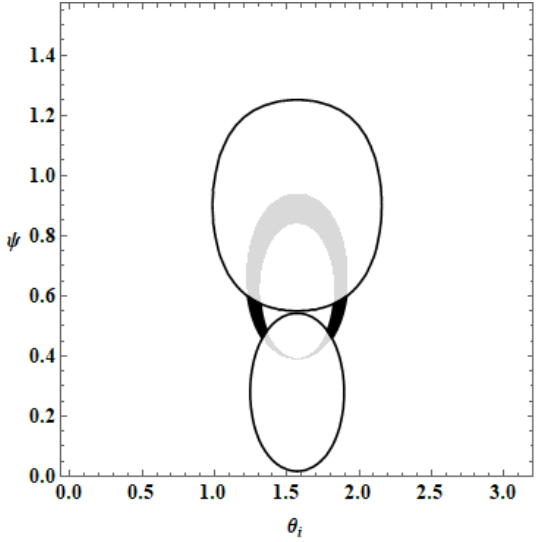

**Figure 9.** A contour plot of the tensor polarization model calculation of $|V/L|$ and $|H/L|$ GW strain signal amplitude ratios that fall within the uncertainty range of the pre-merger observations [See Eq.(22).], plotted as a function of $\theta_i$ and $\psi$, both in radians. Source location: 45° S, 73° W. The solid line contours bound the $|V/L|$ ratio range. The shaded (gray) area indicates the $|H/L|$ ratio range. The solid black regions show the overlap area where the two ratio ranges are matched jointly. The white areas are regions where the calculated $|H/L|$ ratios are outside the specified uncertainty range.

We will use contour plots to display the range of angles for which the predicted amplitude ratios match the GW170814 relative amplitude data. In those figures, we assume that the Hanford/Livingston amplitude ratio falls in the range [1.19, 1.33] and the Virgo/Livingston ratio is in the range [0.86, 1.06]. [See Eq. (22).] As we shall see, the results are not overly sensitive to the range of amplitude ratios as long as that range is generally consistent with the values in Eq. (22).

Figure 9 shows the tensor polarization calculation of the range of orbital inclination and polarization orientation angles in which a tensor polarization model is consistent with the range of observed $|H/L|$ and $|V/L|$ amplitude ratios during the pre-merger inspiral. In this case, the region outside the solid curves indicates the range of $\theta_i$ and $\psi$ that gives ratios consistent with the $|V/L|$ amplitude ratio measurements (taking into account the range of uncertainty). The shaded (gray) region indicates the angular range that gives $|H/L|$ amplitude ratios consistent with the observations. We see that there is a range of angles for which the ratio ranges overlap, indicated by the solid black regions. Thus, we conclude that the tensor polarization model can account for the observations for a modest range of $\theta_i$ and $\psi$. As mentioned previously, this result should not be surprising since the LIGO-Virgo analysis assumed pure tensor polarization and the source location used in figure 9 is the LIGO-Virgo maximum a posteriori sky location. We note in passing that the solid black areas in figure 9 indicate the "support" for the joint likelihood functions to be introduced in section 9. Let us now turn to the vector model predictions.

*6.2    Vector polarization model*

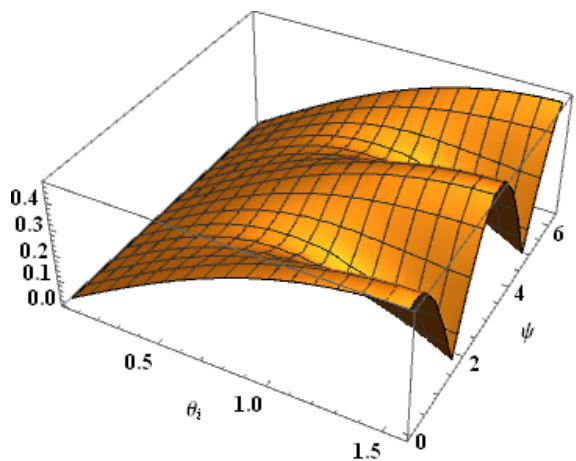

**Figure 10.** Vector polarization model predictions for the Livingston GW strain signal amplitude (arbitrary units) as a function of the orbital angle of inclination $\theta_i$ and the polarization orientation angle $\psi$, both in radians. Source location: 45° S, 73° W. The amplitude is symmetric about the $\theta_i = \pi/2$ line (not shown).

Figure 10 shows a plot of the Livingston observatory GW strain signal amplitude as predicted by the vector polarization model described in section 3.2. Although the absolute amplitude of the signal will depend on the particular vector gravity theory, the geometric part given in section 3.2. will be universal. Hence, the results shown in figure 10 are quite generic. Note that for small values of the orbital inclination angle $\theta_i$, the amplitude is largely independent of the polarization orientation angle $\psi$: in this range the predicted GW emission is approximately circularly polarized; hence, there is no significant polarization orientation angle dependence. Note also that the amplitude goes to zero as $\theta_i \to 0$: in a vector polarization model there is no gravitational wave emission perpendicular to the plane of the binary orbit, in contrast to the tensor polarization model for which the power emission is maximum in that direction [See figure 7.]. So, in a pure vector polarization model, there is an observational bias for edge-on orbits.

As $\theta_i \to \pi/2$, the vector polarization "x-mode" amplitude goes to zero, and the emission is almost completely linearly polarized. In that case, the amplitude is strongly dependent on $\psi$ and repeats every $\pi$ radians, characteristic of vector linear polarization (like E&M linear polarization). The Hanford and Virgo plots are similar.

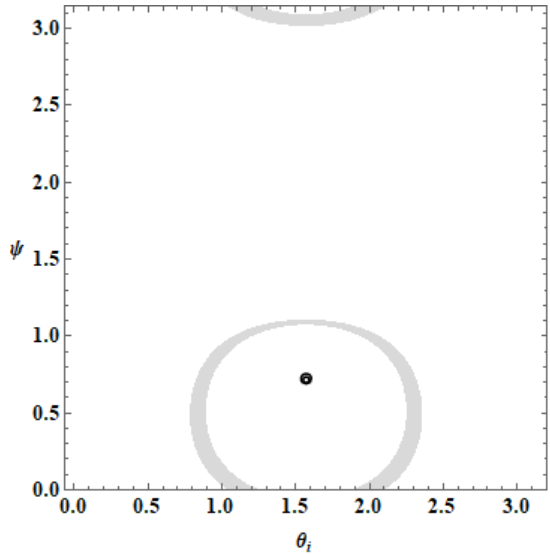

**Figure 11.** A contour plot showing the vector polarization model calculation of the Virgo/Livingston and Hanford/Livingston GW strain signal amplitude ratios that fall within the range of the pre-merger inspiral relative amplitude observations (taking into account the uncertainty range) plotted as a function of $\theta_i$ and $\psi$, both in radians. Source location: 45° S, 73° W. Solid line contours indicate the (very small) angular range that gives $|V/L|$ amplitude ratios consistent with the observations. The shaded (gray) area indicates the $|H/L|$ ratio range. The white areas are regions for which the calculated $|H/L|$ ratios are outside the range of uncertainty.

For this source location, there are no regions for which both ratio ranges are satisfied jointly by the vector polarization model.

Figure 11 displays a contour plot of the vector polarization relative amplitude predictions. We see that the region near $\theta_i \approx \pi/2$ and $\psi \approx 0.7$, for which the vector model predictions agree with the $|V/L|$ amplitude ratio, is very small and that there is no overlap with the region consistent with $|H/L|$ ratio range (gray shading). Hence, we conclude that the vector polarization model cannot account for the observed pre-merger amplitude ratios for the LIGO-Virgo maximum a posteriori GW source location ($45°$ S, $73°$ W).

## 7. Changing the GW source sky location

We now consider how the amplitude ratio predictions vary if we change the GW source sky location. As noted previously, there is a modest range of source locations consistent with the uncertainties in the wave-front arrival times at the detectors. After exploring several combinations of source latitude and longitude, we found a band of source locations (bounded by the dashed line in figure 3) for which a vector polarization model gives results in agreement with the observed amplitude ratios.

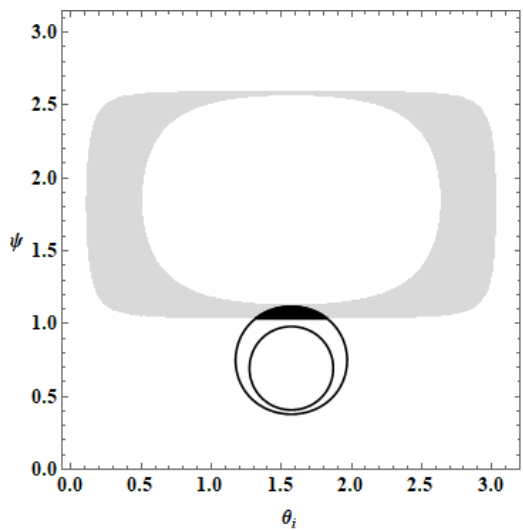

**Figure 12.** A contour plot indicating the vector polarization model calculation of the Virgo/Livingston and Hanford/Livingston GW strain signal amplitude ratios that fall within the uncertainty range of the pre-merger observations plotted as a function of $\theta_i$ and $\psi$, both in radians. Source location: 55°S, 78°W. The region between the solid line contours indicates the $|V/L|$ ratio range. The shaded (gray) area indicates the $|H/L|$ ratio range. The black region is the joint overlap area.

Figure 12 shows the predicted amplitude ratio results for the vector polarization model for a source location 55° S, 78° W. The region between the two solid line contours is consistent with the observed $|V/L|$ amplitude ratios and the shaded (gray) region indicates consistency with the $|H/L|$ ratios. Note that we now have an overlap region shown in black, near $\theta_i \approx \pi/2$, $\psi \approx 1.0$, both in radians. The pattern repeats for $\psi \rightarrow \psi + \pi$ (not shown). Hence, we conclude that for this source location, vector polarization is consistent with the observed amplitude ratios.

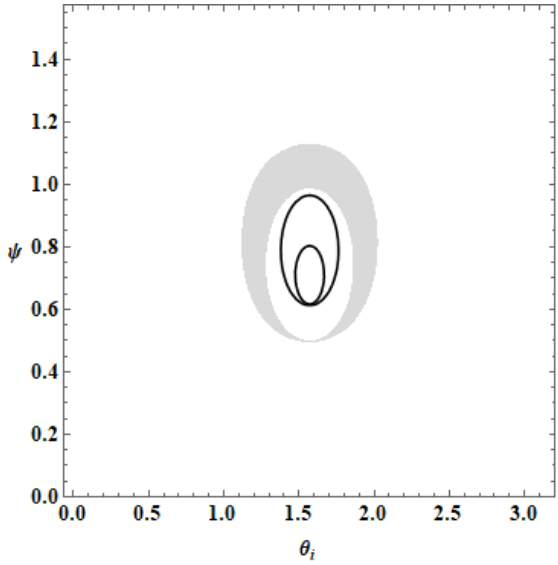

**Figure 13.** A contour plot indicating the tensor polarization model calculations for the $|H/L|$ and $|V/L|$ GW strain signal amplitude ratios that agree with the pre-merger observations [Eq. (22)] plotted as a function of $\theta_i$ and $\psi$, both in radians. Source location: 55° S, 78° W (the same as figure 12). The region bounded by the solid line contours indicates the angular range that gives $|V/L|$ ratios consistent with the observations. The shaded (gray) area indicates the $|H/L|$ ratio range.

Figure 13 shows analogous results for the tensor polarization model. We see that for the specified source location (the same as the one used in figure 12) there is no range of angles for which the tensor polarization model predictions agree with the observed pre-merger amplitude ratios.

Exploring source latitude and longitude combinations, we find that the tensor polarization model gives amplitude ratios consistent with the observations and their uncertainties in the region bounded by the black dotted line in figure 3. The range continues to the left beyond the gnomonic projection region. The analogous region for the vector polarization model is to the right of and below the dashed black line in figure 3. That region continues to the right beyond the gnomonic projection area.

We conclude that the vector polarization model can account for the observed pre-merger amplitude ratios for GW source locations within a range that overlaps and extends beyond the LIGO-Virgo GW170814 90% credible region, but only for relatively small ranges of $\theta_i$ and $\psi$. Note that for $|H/L|>1$, the orbital inclination angle is likely to be close to $\pi/2$ for both the tensor and vector models and hence the variation of the ratios with the polarization orientation angle $\psi$ cannot be ignored.

The results described above indicate that both tensor and vector polarization models can account for the signal amplitude ratios among the three observatories over modest ranges of GW source locations. Thus, we conclude that the observed amplitude ratios do not rule out vector contributions to the GW polarization. Section 9 will use Bayesian model comparison methods to provide quantitative statements about how much the data favor the tensor polarization model over the vector polarization model (or vice versa).

## 8. Phase differences

The observed amplitude ratios are intimately tied to the GW source sky location, the orbital inclination angle and the polarization orientation angle, as shown in the previous section. As we have seen, matching the observed amplitude ratios with those predicted by the tensor polarization model and the vector polarization model allows us to determine, for a specified source location, which model is successful in predicting the correct amplitude ratios. The detectors' relative GW strain signals also depend on the phase differences of the GW strain signals at the observatories due to the orientations of the detectors' arms relative to the GW polarization and propagation directions. We shall call those phases "intrinsic phases" to distinguish them from phases associated with different arrival times at the three detectors. Although our conclusion about the possibility of vector polarization has not made use of any phase data, in principle, the intrinsic phases provide additional information to narrow the possible range of source sky locations, orbital inclination angles, and polarization orientation angles. However, for GW170814, because of the relatively noisy data and the short duration of the GW signal, it is difficult to determine those phase differences with sufficient precision to add significantly to what we have already learned from amplitude ratios alone.

That said, we now examine the intrinsic phase differences between the Hanford and Livingston signals, focusing our attention on just one source location for which the tensor polarization and vector polarization models both account for the observed signal amplitude ratios. To take into account all of the possible signs in Eq. (5), we use the arctan2($y$,$x$) function (the two-argument arctan function). We then fold the phase differences into the range $[-\pi, \pi]$.

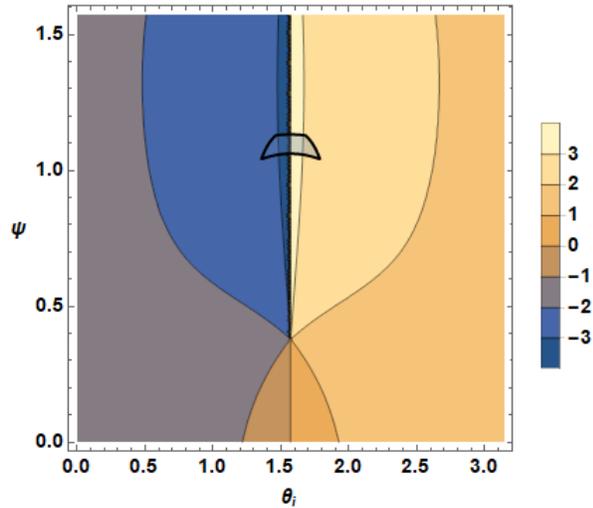

**Figure 14.** A contour plot of the vector model Hanford-Livingston phase difference $(\Phi_H - \Phi_L)$ predictions (in radians) as a function of the orbital inclination angle and the polarization orientation angle (both in radians). Source location: 46°S, 80° W. The small, dark border "wedge" shape region indicates the range of $\theta_i$ and $\psi$ for which the vector model accounts for the signal amplitude ratios. (Compare with figure 12.)

The vector polarization model intrinsic phase difference predictions are shown in figure 14 for the source location $46°S$, $80°W$, which lies in the region bounded by the dashed and dotted lines in figure 3. The pattern repeats for $\psi \to \psi + \pi$ (not shown). Figure 14 indicates a Hanford-Livingston phase shift of about $\pm\pi$ radians over the range of angles for which the vector polarization model accounts for the signal amplitude ratios. That phase shift is consistent with the signals shown in figure 4.

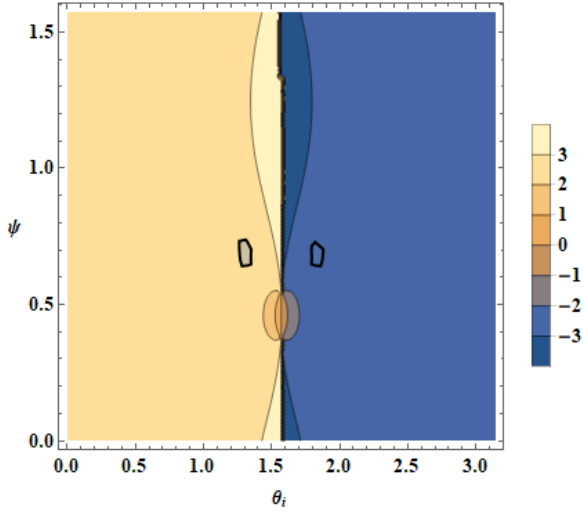

**Figure 15.** A contour plot of the tensor model Hanford-Livingston phase difference $(\Phi_H - \Phi_L)$ predictions (in radians) as a function of the orbital angle of inclination and the polarization orientation angle (both in radians). The small, dark border "wedge" shape regions indicate the range of $\theta_i$ and $\psi$ for which the tensor model accounts for the signal amplitude ratios. Source location: $46°S$, $80°W$, the same as figure 14. The pattern repeats for $\psi \to \psi + \pi/2$ (not shown).

Figure 15 shows the Hanford-Livingston phase difference as predicted by the tensor polarization model. The source location is the same as that used in figure 14. The tensor model predicts a phase difference absolute value a bit smaller than $\pi$.

What do we expect for the Hanford-Livingston phase difference? By design, the Hanford-Livingston interferometer arms are approximately aligned, with $\hat{H}_x \approx -\hat{L}_y$ and $\hat{H}_y \approx \hat{L}_x$. Hence, Eqs. (6) and (8) indicate that we should expect $\pm\pi$ for the $\Phi_H - \Phi_L$ intrinsic phase difference.

However, the alignment is not exact. The angle between $\hat{H}_x$ and $-\hat{L}_y$ is about $24°$ and between $\hat{H}_y$ and $\hat{L}_x$ about $13°$. So, the phase shifts will differ from $\pi$ and are likely to depend significantly on $\theta_i$ and $\psi$ as seen in figures 14 and 15. We saw a similar result with the predicted amplitude ratios shown in figure 8.

Using a waveform based on numerical GR (tensor polarization), the LIGO-Virgo collaboration has shown (see figure 1 of [1]) that the Hanford strain signal data relative to the Livingston strain signal data are consistent with an intrinsic phase of about $\pm\pi$ and an arrival-time delay of about 8 ms during the pre-merger regime. These results indicate that the calculations of the phase shifts presented above are generally consistent with the data.

Given the uncertainties in the observed phase shifts, it is difficult to draw definitive conclusions from the comparison of the predicted intrinsic phase shifts with the observations. We have seen that the phase differences in both the tensor and vector models, vary significantly with $\theta_i$ and $\psi$. In principle, detailed analysis of the phase shifts could lead to further restrictions on the range of orbital inclination angles and polarization orientation angles that are consistent with the observations.

## 9. Bayes factors

We now compute the Bayes factor [42,43,48,49] for the tensor and vector polarization models to provide a quantitative measure of how well the GW170814 data support a tensor polarization model over a vector polarization model (or vice versa). In general terms, the Bayes factor is calculated from the posterior distribution $p(\Theta | y)$ of the model parameters $\Theta$ given data $y$:

$$p^{(T,V)}(\Theta | y) \propto L^{(T,V)}(y | \Theta) p^{(T,V)}(\Theta) , \qquad (23)$$

where we have added superscripts to indicate the tensor polarization model (*T*) or the vector polarization model (*V*). $L^{(T,V)}(y | \Theta)$ is the likelihood of the data given the parameters. $p^{(T,V)}(\Theta)$ is the so-called prior distribution for the model parameters: any prior information we have (or assume) about the distribution of those parameters before we take the data into account. That prior information may be different for different models. It is well known [42,43], that the resulting Bayes factor may depend sensitively on the prior distributions; so, being transparent about the choice of priors is important. Since we will be calculating ratios, the normalization of the posterior distributions will drop out since, by construction, it is the same for both models.

The Bayes factor for a specified pair of models is found by integrating ("marginalizing") the posterior distributions over the parameters and calculating the ratio of the results. (The integrated posterior distribution is sometimes called the "evidence.") We write the Bayes factor $B^{(T/V)}$ for the tensor polarization model relative to the vector polarization model as

$$B^{(T/V)} = \frac{\int L^{(T)}(y|\Theta)p^{(T)}(\Theta)d\Theta}{\int L^{(V)}(y|\Theta)p^{(V)}(\Theta)d\Theta}. \tag{24}$$

In the LIGO-Virgo Collaboration's data analysis, Bayesian methods are used to infer the values of about 15 parameters with a likelihood function built from the Fourier transform of various models ("templates") of the GW wave form and the Fourier transform of the strain time series (GW signal plus noise). The statistical distribution of each inferred parameter and the maximum likelihood value of that parameter are calculated by integrating the posterior distributions over all the other parameters. In our approach, we take advantage of the fact that once the GW source sky location is specified, the only relevant parameters for determining GW polarization are the relative (complex) strain amplitudes among the observatories, the orbital inclination angle and the polarization orientation angle. Consequently, there is a substantial reduction in the dimensionality of the relevant parameter space. In essence, we are integrating over any parameters that are not relevant to extracting GW polarization information from the measured amplitude ratios and phase differences. Since the priors for the irrelevant variables can be taken to be the same for the two models, we may simply ignore those irrelevant parameters according to standard Bayesian inference methods.

We employ that reduction in parameter space to formulate directly the likelihood functions in terms of those remaining parameters. For our analysis, the model parameters are given by $\Theta = \{\theta_i, \psi, \theta_{lat}, \phi_{long}\}$. The data are the amplitude ratios and phase differences determined by the three methods as described in section 5, though it turns out the amplitude ratios are more important than the phase differences except for the approximately $\pm\pi$ phase difference between the Hanford and Livingston GW signals.

Given a set of data $\{y\}$, the joint likelihood functions $L^{(T,V)}(\{y\},\Theta)$ for the $|H/L|$ and $|V/L|$ ratios are expressed by

$$L^{(T,V)}(\{y\},\Theta) = exp\left[-N_L^{(T,V)}(\{y\},\Theta)\right], \tag{25}$$

where $N_L^{(T,V)}$ is the negative of the natural log of the likelihood function:

$$N_L^{(T,V)} = \sum_j \left\{ \frac{\left[|H/L|_j - r_{HL}^{(T,V)}(\Theta)\right]^2}{2\sigma_{HLj}^2} + \frac{\left[|V/L|_j - r_{VL}^{(T,V)}(\Theta)\right]^2}{2\sigma_{VLj}^2} \right\}. \quad (26)$$

In Eq. (26), $|H/L|_j$ is the $j$th ratio data value, $\sigma_{HLj}^2$ is its variance, and $r_{HL}^{(T,V)}(\Theta)$ is the predicted ratio, with analogous expressions for $|V/L|$.

To evaluate the posterior distribution, we need to specify the prior information. We choose a $\theta_i$ distribution uniform on $[0,\pi]$ and a $\psi$ distribution uniform on $[0, \pi/2]$ for the tensor polarization model and uniform on $[0,\pi]$ for the vector polarization model. Recall that the $\psi$ dependence of the tensor polarization amplitudes has a periodicity of $\pi/2$ while the vector polarization has a periodicity of $\pi$. See figures 7 and 10. Integrating the tensor model likelihood over the $\psi \in [0,\pi]$ range would double its marginalized value even though its predictive power is unchanged. Hence, we restrict the tensor polarization model to the more restricted angular range. If that effect is ignored, the Bayes factor for the tensor polarization model versus the vector model would increase by a factor of two, which, as we shall see, has no significant effect on the model comparison results.

The priors for the source latitude and longitude are expressed by the range of sky locations to be included in the integral of the posterior distribution. Specifying and justifying those ranges are important [23]. To illustrate the effects of choosing different source latitude and longitude ranges, we will evaluate the Bayes factors for four regions: (1) the range of latitude and longitude corresponding to the gnomonic projection in figure 3; (2) the range of latitude and longitude corresponding to the region that satisfies the GW arrival-time delays (and their uncertainties) (the region in figure 3 bounded by the gray lines); (3) the range associated with the LIGO-Virgo 90% credible range of sky locations (the lobe-shaped region in figure 3); and (4) the lower-right region of the gnomonic projection in figure 3 ($\theta_{lat} < -50°$ and $\phi_{long} < -73°$). The results for the Bayes factors are given in Table I.

Table I. The tensor versus vector Bayes factors for the four integration regions (see text) and two different values for the one-sigma uncertainty in $|V/L|$.

| Region | $B^{(T/V)}, \sigma_{VL} = 0.1$ | $B^{(T/V)}, \sigma_{VL} = 0.2$ |
|--------|-------------------------------|-------------------------------|
| 1 | 1.5 | 1.9 |
| 2 | 1.0 | 0.7 |
| 3 | 84 | 36 |
| 4 | 0.1 | 0.2 |

The integrations were carried out for two different values of the uncertainty associated with $|V/L|$. As mentioned previously, the Virgo data are particularly noisy, so using a large uncertainty generally gives a less decisive Bayes factor. The uncertainty in the individual Bayes factors is about 20%, determined by varying the numerical integration parameters and by making modest adjustments in the boundaries of the source latitude and longitude integration regions.

The important lesson to be drawn from the results in Table I is that the Bayes factor depends sensitively on the assumed prior for the source sky location and less sensitively on the assumed uncertainties. For Regions 1 and 2, the Bayes factors tell us that the data provide equal support to vector and tensor polarization models. Only in Region 3 do we find (unsurprisingly) that the data favor tensor polarization since the 90% credible region was found assuming tensor polarization. Region 4 was chosen intentionally to favor vector polarization.

We argue that using Region 2, which makes use of our prior knowledge of the (largely model-independent) arrival-time delays and their uncertainties, is the preferred choice for the source location prior. The critical conclusion is that the GW170814 data give roughly equal support to both the tensor and vector polarization models and hence cannot rule out GW vector polarization.

## 10. Discussion

In this paper we have used a geometric analysis employing specified GW source locations to explore the ranges of orbital inclination angles and polarization orientation angles that give results consistent with the observed GW170814 signal amplitude ratios found in section 5. We then used those source locations to examine intrinsic phase shifts between the signals from the three observatories as functions of orbital inclination and polarization orientation angles. Finally, we integrated over the four model parameters $\Theta = \{\theta_i, \psi, \theta_{lat}, \phi_{long}\}$ to find the Bayes factor for the tensor model relative to the vector model. Our analysis focused on the GW170814 signal during the pre-merger inspiral stage of the binary motion. That focus allows us to ignore the potential complications that arise during the merger and ring-down stages [19,36] due to strong spacetime curvature and the concomitant scattering of the GWs in the source near-field region, leading to possible changes in the relative amplitudes and phases of the GW polarization modes. We could, in principle, employ the techniques described in this paper to find amplitude ratios during the merger and ring-down stages to extract polarization information for those stages. However, those stages include only a few GW cycles, and we anticipate that the polarization amplitudes (and their relative phases) might change rapidly during merger and ring-down.

The analysis provided here complements the full Bayesian analysis provided by the LIGO-Virgo collaboration by giving a more transparent treatment of the interplay among GW source sky location, the signal amplitude ratios and the GW polarization modes. In particular, the analysis establishes the importance of considering the full range of orbital inclination and polarization orientation angles in determining the possible contributions of tensor and vector polarization modes, and the need to specify explicitly the range of source sky locations over which the posterior distributions are marginalized for Bayesian model comparisons.

Two recent papers [15,28] also emphasize the importance of considering explicitly the full range of orbital inclination and polarization orientation angles and the sensitivity of conclusions about polarization modes to the source sky location parameters. Ref. [15] focuses on determining a possible mix of polarization modes in signals from coalescing binaries as observed by a network of GW detectors. The emphasis is on how various models of mixed polarization modes affect the uncertainties in the parameters deduced from the analysis. The authors argue that under favorable circumstances, the GW data from three detectors are sufficient to draw conclusions about the mix of polarization modes.

Ref. [28] uses a null stream analysis with a model of GW vector polarization with a time-dependent polarization orientation angle to show that vector GW signals can mimic tensor GW signals over a range of source sky locations. The polarization orientation angle would vary in time if the binary source exhibited significant spin-orbit and spin-spin interactions, as mentioned previously. However, the LIGO-Virgo analysis of GW170814 data found no sign of spin interactions.

We note that tests of the polarization content of GWs can also be carried out with continuous wave GWs (if observed) [6,27]. Here "continuous" means that the GW signal duration is a significant fraction of a day, so the angles between the detector arms and the GW propagation direction and the polarization orientation change significantly during the observations. The most likely sources for continuous GWs whose frequencies fall within the LIGO-Virgo detectors' frequency bands are pulsars. The recent search [6] for such signals yielded a null result. Also, a recent search for tensor, vector, and scalar polarizations in the stochastic GW background [50] yielded a null result.

We have seen that the results of our analysis support the three previously mentioned premises: (1) Given a specified GW source location, one can calculate the predicted relative amplitudes for the GW waveform detected by the three LIGO-Virgo observatories based on a purely geometric method, but the results depend on the orbital inclination angle and the polarization orientation angle. (2) The relative signal amplitudes among the three observatories can be extracted from the GW170814 data presented in [1] with uncertainties small enough to make significant comparisons with the predictions for vector and tensor polarization models. (3) The comparison between the observed and predicted relative amplitudes indicates that both tensor and vector polarization models can account for the relative amplitude observations for a range of source locations consistent with the results presented in [1].

In principle, as mentioned in section 1, GWs could have contributions from all possible polarization modes: tensor, vector and scalar. The contributions for GWs emitted by orbiting binaries depend on the detailed dynamical theories that give rise to those polarization modes. See [16] for a pure vector theory of gravity that gives the same quadrupole energy loss expression as GR for the inspiral dynamics. Ref. [51] provides a survey of several gravity models and their GW polarization properties.

Ref. [1] claims that the LIGO-Virgo analysis of GW170814 polarization favors tensor over vector polarization with a Bayes factor of 200 (later reduced to 30 [23]). We have shown in section 9 that only by restricting the range of sky locations to the LIGO-Virgo 90% credible sky location region do we get a Bayes factor close to 30. However, lacking the details of the LIGO-Virgo analysis (for example, the range of source locations and the range of orbital inclination angles and polarization orientation angles examined), it is not possible to tell if the LIGO-Virgo result is consistent with the results presented in this paper. Examination of footnote 19 in [22] suggests that the disagreement is due to different choices of priors for the source sky locations.

We have demonstrated how knowledge of the source sky location (or lack thereof) affects our ability to sort out possible GW polarization modes [41]. It turns out that for the range of physically reasonable GW1701814 source locations, the three LIGO-Virgo observatories do not provide high discrimination among polarization modes. See figure 2 of [41]. Nonetheless, the analysis presented in this paper shows explicitly that the GW170814 data cannot preclude vector polarization for the range of source locations discussed in the previous section. Consequently, for definitive tests of GW polarization from GW burst events, we need to look at binary neutron star or neutron star-black hole coalescences, whose locations can be precisely determined via the concurrent electromagnetic emissions. In particular, the binary neutron star GW170817 event [33] allows for a more definitive determination of GW polarization content because the source sky location is well-established through the concomitant electromagnetic emissions of the neutron star merger. The details of the polarization analysis of GW170817 are given in [35], which provides convincing evidence that the GW170817 data strongly favor vector polarization over tensor polarization. We anticipate that observations of future binary neutron star or neutron star-black hole coalescences will provide further evidence to confirm that conclusion.

**Acknowledgements**

The author thanks Nicolas Arnaud of the EGO/Virgo Outreach Team for directions to the gravitational wave detector location and orientation information. C. Mead and T. A. Moore provided several helpful discussions and comments. A. Svidzinsky, A. Weinstein, and M. Isi provided detailed comments and advice on drafts of this paper.

This research has made use of data, software and/or web tools obtained from the LIGO Open Science Center (https://www.gw-openscience.org), a service of LIGO Laboratory, the LIGO Scientific Collaboration and the Virgo Collaboration. LIGO is funded by the U.S. National Science Foundation. Virgo is funded by the French Centre National de Recherche Scientifique (CNRS), the Italian Istituto Nazionale della Fisica Nucleare (INFN) and the Dutch Nikhef, with contributions by Polish and Hungarian institutes.